%% file: SungPoorYu.tex
\documentclass{article}

\def\ignore#1{}

\input macros

\input setup
\input labelfig.tex

\usepackage{spconf,amsmath,epsfig,epsf,psfrag,amssymb,amsfonts,latexsym, amsmath,color}
\usepackage{verbatim}
\usepackage[mathscr]{eucal}

\newcommand{\beq}{\begin{equation}}
\newcommand{\eeq}{\end{equation}}

\def\defeq{\stackrel{\Delta}{=}}
\def\Ebb{{\mathbb E}}

\definecolor{bgrd}{rgb}{1,1,1}
\definecolor{grey}{rgb}{0.9,0.9,0.6}
\definecolor{gray}{rgb}{0.5,0.5,0.5}

\def\L{{\cal L}}

\title{Large Deviations Analysis for the Detection of 2D Hidden Gauss-Markov Random Fields Using Sensor Networks}
\name{Youngchul Sung\sthanks{{\scriptsize Y. Sung and H. Yu are
with the Dept. of Electrical Engineering, Korea Advanced Institute
of Science and
  Technology (KAIST), Daejeon 305-701, South Korea.  Email:ysung@ee.kaist.ac.kr and hjyu@stein.kaist.ac.kr. H. V. Poor is with the Dept. of Electrical Engineering,
  Princeton University, Princeton, NJ 08544. Email: poor@princeton.edu. The work of Y. Sung was  supported in
part by Brain Korea 21 Project, the School of Information
Technology, KAIST. The work of H. V. Poor was supported in part by the U. S. National
Science Foundation under Grants ANI-03-38807  and CNS-06-25637.}}, H. Vincent Poor and Heejung Yu }
\address{}

\begin{document}
\maketitle

\ninept
{\footnotesize
\begin{abstract}
The detection of hidden two-dimensional Gauss-Markov random fields
using sensor networks is considered.  Under a conditional
autoregressive  model, the error exponent for the Neyman-Pearson
detector satisfying a fixed level constraint is obtained using the
large deviations principle. For a symmetric first order
autoregressive model, the error exponent is given explicitly
 in terms of  the SNR and an edge dependence factor (field correlation).
 The behavior of the error exponent as a function of
 correlation strength is seen to divide into two regions depending on
the value of the SNR. At high SNR, uncorrelated observations maximize the error
exponent for a given SNR, whereas there is non-zero optimal
correlation at low SNR. Based on the error exponent, the energy
efficiency (defined as the ratio of the total information gathered to the
total energy required)  of {\em ad hoc} sensor network for  detection
 is examined for two sensor deployment models: an infinite
area model and and infinite density model. For a fixed sensor density,
the energy efficiency diminishes to zero at  rate
$O(\mbox{area}^{-1/2})$ as the area is increased. On the other
hand, non-zero efficiency is possible for increasing density
depending on the behavior of the physical correlation as a function of the link
length.
\end{abstract}
}

{\footnotesize \textbf{\textit{Index Terms-}}  Neyman-Pearson
detection, error exponent, GMRF}

\vspace{-1.2em}
\section{Introduction}

\vspace{-1em} Consider the design of a sensor network for the
detection of a correlated stochastic signal in a fixed area.
Many questions arise in such a design: How do the field correlation
and measurement signal-to-noise (SNR) affect the detection
performance? What is the optimal sensor density, i.e., the number
of nodes per unit area?  What is the information and energy
trade-off in such a sensor network with {\em ad hoc} routing? To
address these issues, several studies based on one-dimensional (1D)
spatial signal models have been conducted (see, e.g.,
\cite{Sung&Tong&Poor:05IPSN} and \cite{Chamberland&Veeravalli:06IT}).
However, there is an important difference between 1D signal models
and actual spatial signals. Suppose that we take observations
from sensors placed equidistantly along a line transect laid over
a given area. The observations may then be viewed as samples generated by a
one-dimensional process and the results from time series analysis
could be applied to investigate their statistical properties.
However, there is no real notion of `signal flow' or dependence
direction along the transect as there is in a more traditionally obtained
 time series. For samples from
sensors deployed over a two-dimensional (2D) area, it is necessary
to consider the signal dependence in all direction in the plane,
and  as a consequence, answering  the above questions becomes more difficult.

To address the above questions in a 2D setting, in this paper, we
consider the detection of 2D Gauss-Markov random fields (GMRFs)
using noisy observations. In particular we consider Sensors  $ij$
located on a 2D lattice ${\mathcal I}$.  On denoting the (noisy)
measurements of
 Sensor $ij$ as $Y_{ij}$ and adopting a
Neyman-Pearson formulation, we can model the detection problem via
 null and alternative hypotheses given by {\footnotesize
\begin{equation} \label{eq:hypothesis2d}
\Hc_0:  Y_{ij} = W_{ij} , ij \in {\cal I}   \ \ {\rm vs.}~~~~~~~ \Hc_1:  Y_{ij} = X_{ij}+ W_{ij}, ij \in {\cal I},
\end{equation}}
where  $\{W_{ij}\}$ represents independent and identically
distributed (i.i.d.) $\Nc(0,\sigma^2)$ noise with a known variance
$\sigma^2$, and $\{X_{ij}\}$ is a stationary GMRF on the 2D
lattice ${\mathcal I}$ independent of the measurement noise
$\{W_{ij}\}$. Thus, the observation samples form a 2D hidden GMRF
under $\Hc_1$.
\begin{figure}[htbp]
\centerline{
    \begin{psfrags}
    \psfrag{ij}[l]{{\scriptsize $(i,j)$}}
    \psfrag{xij}[c]{{\scriptsize $X_{ij}$}}
    \psfrag{wij}[l]{{\scriptsize $W_{ij}$}}
    \psfrag{yij}[c]{{\scriptsize $Y_{ij}$}}
    \psfrag{Nij}[c]{{\scriptsize Sensor $ij$}}
    \psfrag{r}[c]{{\scriptsize $r$}}
    \scalefig{0.30}\epsfbox{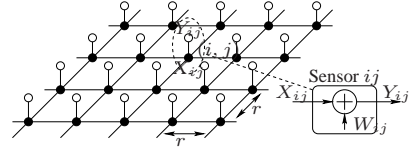}
    \end{psfrags}
} \caption{Sensors on a 2D Lattice ${\mathcal I}$: Hidden Markov
Structure} \label{fig:2dHGMRF}
\end{figure}

%%%%%%%%%%%%%%%%%%%%%%%%%%%%%%%%%%%%%%%%%%%%%%%%%%%%%%%%%%%%%%%%%%
\vspace{-1.5em}
\subsection{Summary of Results}
\vspace{-0.5em}

The exact error probability of the  detection of the
Neyman-Pearson test is not available in closed-form in the general
correlated case, including the hypotheses (\ref{eq:hypothesis2d}).
Hence, we invoke the the large deviations principle and use the
{\em error exponent} of the detection probability (or, more
conveniently, its complement, the miss probability) as an
alternative performance measure. For a fixed false-alarm level,
the miss probability $P_M$ decays exponentially as the sample size
$n$ increases, and the {error exponent} is defined as the rate of
exponential decay, i.e., {\footnotesize
\begin{equation}
\Kc \defeq \lim_{n\rightarrow\infty} -\frac{1}{n} \log P_M
\end{equation}}
under the given  constraint (i.e., the false alarm probability
$P_F \le \alpha$).  The error exponent is a good performance
criterion in the large sample regime since it allows the designer to estimate
the number of samples required for a given detection
performance. Hence,  efficient design  can be examined through
the error exponent for large scale sensor networks.

Here, we adopt the {\em  conditional autoregression (CAR) model} for the
signal, and derive a closed-form expression for the error exponent
 $\Kc$
 of the miss probability (which is independent of $\alpha$) in the spectral domain.
 We do so by
 exploiting the spectral structure of the CAR signal and the relationship between the eigenvalues of the block circulant
 approximation to a block Toeplitz matrix describing the 2D correlation
 structure.  In particular, it is shown that the error exponent for the detection
 of 2D hidden GMRF is an extension of that in the 1D case obtained by
 Sung et al.\cite{Sung&Tong&Poor:06IT}.  As in the 1D case, it is
 shown that  i.i.d. (and, thus, uncorrelated) observations maximize the error exponent
 for a given SNR when the SNR is high. On the other hand, there is an
 optimal non-zero degree of correlation at low SNR. Interestingly, it is
 seen that there is a discontinuity in the optimal correlation
 strength as a function of SNR.  In the
 perfectly correlated case, the error exponent is zero as expected.
 For the error exponent as a
 function of SNR, we will show that the error exponent increases as
 $\log \mbox{SNR}$ for a given correlation strength at high SNR.

We consider two asymptotic regimes modelling the sensor deployment
in 2D: an infinite area model with a fixed density and an infinite
density model with a fixed area. Applying the results, we obtain
the asymptotic behavior of the energy efficiency, defined as the
ratio of the total information gathered to the required energy to obtain
information from the area for an ad hoc network with minimum hop
routing to the fusion center. For the infinite area model, the
energy efficiency decays to zero with rate
$O({{\mbox{area}}^{-1/2}} )$ as we increase the coverage area. For
the infinite density model, on the other hand, a non-zero
efficiency is possible if the decay rate of the error exponent
$\Kc(\mbox{density})$ as a function of density is slower than
$O(\mbox{density}^{(1-\delta)/2})$, where $\delta$ is the
propagation constant $\delta \ge 2$.

\vspace{-1.em}
%%%%%%%%%%%%%%%%%%%%%%%%%%%%%%%%%
\subsection{Related Work}
%%%%%%%%%%%%%%%%%%%%%%%%%%%%%%%%%
\vspace{-0.5em}

The detection of Gauss-Markov processes in Gaussian noise is a
classical problem. See \cite{Kailath&Poor:98IT} and references
therein. However, most work in this area considers only  1D signals or time series. A
closed-form error exponent was obtained and its properties were
investigated for 1D hidden Gauss-Markov random processes
\cite{Sung&Tong&Poor:06IT}. Large deviations analyses were used
to examine the  issues of optimal sensor density  and optimal sampling
were examined with a 1D signal model in
\cite{Sung&Tong&Poor:05IPSN} and \cite{Chamberland&Veeravalli:06IT}.

 An error
exponent was obtained for the detection of 2D GMRFs in
\cite{Anandkumar&Tong&Swami:07ICASSP}, where the sensors are
located randomly and the Markov graph is based on the nearest
neighbor  dependency enabling a loop-free graph and further
analysis. In this work, however, the measurement noise was not
captured. Our work here focuses on the error exponent for the
detection of 2D {\it hidden} GMRF on a 2D infinite lattice, which
allows for the consideration of measurement noise.  In particular we
examine the above  CAR
model and investigate of the detection performance with
respect (w.r.t.) to various design parameters such as correlation strength,
measurement SNR, sensor density and area.

%%%%%%%%%%%%%%%%%%%%%%%%%%%%%%%%%%%%%%%%%%%%%%%%%%%%%%%%%%%%%%%%%%%%%%%
\vspace{-1.0em}
\section{Data Model}
\label{sec:systemmodel}

\vspace{-0.5em}
\begin{definition}[GMRF \cite{Rue&Held:book}]\label{def:GMRF}
A random vector $\Xbf=(X_1,X_2,\cdots,X_n)$ $\in {\mathbb R}^n$ is
a Gauss-Markov random field  w.r.t. a labelled graph ${\mathcal
G}=({\mathcal \nu},{\mathcal E})$ with mean $\mubf$ and precision
matrix $\Qbf
>0$, if its probability density function is given by {\footnotesize
\begin{equation}
p(\Xbf) = (2\pi)^{-n/2}|\Qbf|^{1/2}\exp\left( - \frac{1}{2}
(\Xbf-\mubf)^T \Qbf (\Xbf-\mubf) \right),
\end{equation}}
and $Q_{lm} \ne 0 \Longleftrightarrow \{l,m\} \in {\mathcal
E}~\mbox{for all}~ l \ne m$.  Here, ${\mathcal \nu}$ is  the set of all nodes
$\{1,2,\cdots, n\}$ and ${\mathcal E}$ is the set
of edges connecting pairs of nodes, which represent the conditional
dependence structure.
\end{definition}

\vspace{-0.7em} Note that the mean and the precision matrix fully
characterize a GMRF. Note also that the covariance matrix
$\Qbf^{-1}$ is completely dense in general while the precision
matrix $\Qbf$ has nonzero elements $Q_{lm}$ only when there is an
edge between nodes $l$ and $m$ in the Markov random field. Hence,
when the graph is not fully connected, the precision matrix is
sparse. The 2D indexing scheme $(i,j)$ can be properly converted
to an 1D scheme to apply Definition \ref{def:GMRF}. From here on,
we use the 2D indexing scheme for convenience.

\vspace{-0.5em}
\begin{definition}[Stationarity]
A 2D GMRF  on 2D doubly infinite lattice ${\mathcal I}_\infty$ is
said to be {stationary} if the mean vector is constant and $Cov(
X_{ij}, X_{i^\prime j^\prime}) \defeq  \Ebb \{X_{ij} X_{i^\prime
j^\prime}\}= c(i-i^\prime, j-j^\prime)$ ~for some function
$c(\cdot, \cdot)$.
\end{definition}
\vspace{-0.5em} For a 2D stationary GMRF $\{X_{ij}\}$, the
covariance $\{\gamma_{ij}\}$ is defined as
\begin{equation}
\gamma_{ij} = \Ebb \{ X_{i^\prime j^\prime} X_{i^\prime+i,
j^\prime +j}\} =\Ebb \{ X_{00} X_{ij}\},
\end{equation}
which does not depend on $i^\prime$ or $j^\prime$ due to the
stationarity. Further, the spectral density function of a
zero-mean and stationary Gaussian process $\{X_{ij}\}$ on
${\mathcal I}_\infty$ with covariance $\gamma_{ij}$ is defined as
\begin{equation}  \label{eq:2DDTFT}
f(\omega_1,\omega_2) =  \frac{1}{4\pi^2}\sum_{ij \in {\mathcal
I}_\infty} \gamma_{ij} \exp(-\iota(i\omega_1 + j\omega_2) ),
\end{equation}
where $\iota = \sqrt{-1}$ and $(\omega_1,\omega_2) \in
(-\pi,\pi]^2$. Note that this is a 2D extension of the conventional 1D
discrete-time Fourier transform (DTFT).

\vspace{-0.5em}
\begin{definition}[The conditional autoregression (CAR)]
A GMRF can be specified using a set of full conditional normal
distributions with mean and precision: {\footnotesize
\begin{eqnarray}
\Ebb \{ X_{ij}|\Xbf_{-ij}\} &=&  -\frac{1}{\theta_{00}}
\sum_{i^\prime j^\prime \in {\mathcal I}_\infty \ne 00}
\theta_{i^\prime j^\prime} X_{i+i^\prime,j+j^\prime}, \label{eq:condMean2DInf}\\
\mbox{Prec}\{X_{ij}|\Xbf_{-ij}\} &=& \theta_{00} > 0,
\label{eq:condPrec2DInf}
\end{eqnarray}}
where $\Xbf_{-ij}$ denotes the set of all variables except
$X_{ij}$.
\end{definition}\vspace{-0.5em}
It is shown that the GMRF defined by the CAR model
(\ref{eq:condMean2DInf}) - ( \ref{eq:condPrec2DInf}) is a
zero-mean stationary Gaussian process on ${\mathcal I}_\infty$
with the spectral density function \cite{Rue&Held:book}
{\footnotesize
\begin{equation}
f(\omega_1,\omega_2) =  \frac{1}{4\pi^2} \frac{1}{\sum_{ij \in
{\mathcal I}_\infty} \theta_{ij} \exp(-\iota (i\omega_1 +
j\omega_2))}
\end{equation}}
if {\footnotesize
\begin{eqnarray}
&&|\{\theta_{ij} \ne 0\}| < \infty, ~~~~ \theta_{ij} = \theta_{-i,-j}, ~~~~ \theta_{00} >0, \label{eq:CARcond1}\\
&&\{\theta_{ij}\} ~\mbox{is so that}~ f(\omega_1,\omega_2)>0, ~~~
\forall (\omega_1,\omega_2) \in (-\pi,\pi]^2. \label{eq:CARcond4}
\end{eqnarray}}
We assume that the 2D stochastic signal in (\ref{eq:hypothesis2d})
is given by a stationary GMRF defined by the CAR model
(\ref{eq:condMean2DInf}) - (\ref{eq:condPrec2DInf}) and
(\ref{eq:CARcond1}) -  (\ref{eq:CARcond4}). Then, the observation
spectrum under the two hypotheses (\ref{eq:hypothesis2d}) are
given, respectively, by
\[
S_{0}^y(\omega_1,\omega_2) = \frac{\sigma^2}{4\pi^2} \ \ {\rm and} \ \
S_{1}^y(\omega_1, \omega_2) = \frac{\sigma^2}{4\pi^2}   +
f(\omega_1,\omega_2).
\]

%%%%%%%%%%%%%%%%%%%%%%%%%%%%%%%%%%%%%%%%%%%%%%%%%%%%%%%%%%%%%%%%%%%
\vspace{-1.3em}
\section{Performance Measure: Error Exponent}
\vspace{-0.5em}
%%%%%%%%%%%%%%%%%%%%%%%%%%%%%%%%%%%%%%%%%%%%%%%%%%%%%%%%%%%%%%%%%%%

In this section, we investigate the performance of the Neyman-Pearson
detector with level $\alpha \in (0, 1)$ for a 2D CAR signal in
noisy observations.   We obtain the error exponent in the spectral
domain for  this problem by
 exploiting the spectral structure of the CAR signal and the relationship between the eigenvalues of  block circulant and block Toeplitz matrices representing 2D correlation
 structure.

\begin{theorem}[Error Exponent]  Consider Neyman-Pearson detection
between the hypotheses
(\ref{eq:hypothesis2d})  with the model (\ref{eq:condMean2DInf}) -
(\ref{eq:condPrec2DInf}) and with level $\alpha \in (0,1).$ Assuming that conditions
 (\ref{eq:CARcond1} and  \ref{eq:CARcond4}) hold, the error
exponent of the miss probability is independent of $\alpha$ and is
given by {\scriptsize
\begin{eqnarray}
\Kc &=& \frac{1}{4\pi^2} \int_{-\pi}^{\pi} \int_{-\pi}^{\pi}
\biggl( \frac{1}{2}\log
\frac{\sigma^2+4\pi^2f(\omega_1,\omega_2)}{\sigma^2}\nonumber\\
&& ~~~~~~~~~~~~~~~~~~~~~~~~~~~~~~~~~~~ +\frac{1}{2}
\frac{\sigma^2}{\sigma^2+4\pi^2f(\omega_1,\omega_2)} -\frac{1}{2}
\biggr)d\omega_1d\omega_2,\label{eq:errorexponentspectral}\\
&=&\frac{1}{4\pi^2}\int_{-\pi}^{\pi} \int_{-\pi}^{\pi} D(
\Nc(0,S_{0}^y(\omega_1,\omega_2))||\Nc(0,S_{1}^y(\omega_1,\omega_2))
~d\omega_1 d\omega_2,\nonumber
\end{eqnarray}
} where $D(\cdot||\cdot)$ denotes the Kullback-Leibler divergence.
\end{theorem}

\vspace{0.5em} {\em Proof:} $\Kc$ is given by the almost-sure
limit of the asymptotic Kullback-Leibler rate {\scriptsize $\Kc=
\lim_{n\rightarrow\infty} \frac{1}{n} \log
\frac{p_{0,n}}{p_{1,n}}(\ybf_n)$} evaluated under
$p_{0,n}$\cite{Vajda:book}. Using the fact that we have Gaussian
distributions under both hypotheses, we have {\scriptsize
\begin{eqnarray*}
\Kc&=& \lim_{n\rightarrow \infty} \frac{1}{n} \left(
\frac{1}{2}\log \frac{\det(\Sigmabf_{1,n})}{\det(\Sigmabf_{0,n})}
+ \frac{1}{2} \ybf_n^T (\Sigmabf_{1,n}^{-1} -
\Sigmabf_{0,n}^{-1})\ybf_n \right),
\end{eqnarray*}}
Then approximating the  block Toeplitz correlation matrix with a block
circulant matrix and applying the the 2D Grenander-Szeg\"o theorem, we  obtain
the limit of each term as follows. {\scriptsize
\begin{eqnarray*}
\frac{1}{n} \log \det(\Sigmabf_{1,n}) &\rightarrow& \frac{1}{4\pi^2} \int_{-\pi}^{\pi} \int_{-\pi}^{\pi}\log(\sigma^2 + 4\pi^2f(\omega_1,\omega_2))d\omega_1d\omega_2,\\
\frac{1}{n} \log \det(\Sigmabf_{0,n}) &\rightarrow& \log \sigma^2,\\
\frac{1}{n} \ybf_n^T \Sigmabf_{1,n}^{-1} \ybf_n &\rightarrow&  \frac{1}{4\pi^2} \int_{-\pi}^{\pi} \int_{-\pi}^{\pi}\frac{\sigma^2}{\sigma^2+4\pi^2f(\omega_1,\omega_2)}d\omega_1d\omega_2,\\
\frac{1}{n} \ybf_n^T \Sigmabf_{0,n}^{-1} \ybf_n &\rightarrow& 1,
\end{eqnarray*}}
almost surely. $\blacksquare$

\vspace{0.5em} This theorem is a 2D extension of the error exponent of 1D
hidden Gauss-Markov model based on state-space structure obtained
in \cite{Sung&Tong&Poor:06IT}. Intuitively, the error exponent
(\ref{eq:errorexponentspectral})  can be explained using the
frequency binning argument. For each 2D frequency segment
$d\omega_1 d\omega_2$, the spectra are flat, i.e., the signals are
independent and Stein's lemma can be applied for the segment. The
overall Kullback-Leibler divergence is the sum of contributions
from each bin.

%%%%%%%%%%%%%%%%%%%%%%%%%%%%%%%%%%%%%%%%%%%%%%%%%%%%%%%%%%%%%%%%%%
\vspace{-1em}
\subsection{Symmetric First Order Autoregression}

To investigate the behavior of the error exponent as a function of
correlation and SNR, we further consider the symmetric first order
autoregression (SFAR), described by the conditions
\vspace{-0.5em}{\footnotesize
\begin{eqnarray*}
\Ebb \{ X_{ij}|\Xbf_{-ij}\} &=&  \frac{\lambda}{\kappa} (X_{i+1,j}+X_{i-1,j}+X_{i,j+1}+X_{i,j-1}),\\
\mbox{Prec}\{X_{ij}|\Xbf_{-ij}\} &=& \kappa > 0,
\end{eqnarray*}}
where $0 \le \lambda \le \frac{\kappa}{4}$. (This is a sufficient
condition to satisfy (\ref{eq:CARcond1}) - (\ref{eq:CARcond4}).)
Note here that $\theta_{00}=\kappa$ and $\theta_{1,0} =
\theta_{-1,0} = \theta_{0,1} = \theta_{0,-1} = -\lambda$. In this
model, the correlation is symmetric for each set of four
neighboring nodes. The SFAR model is a simple yet meaningful
extension of the 1D Gauss-Markov random process, which has the
conditional causal dependency only on the previous sample. Here in
the 2D case we have four neighboring nodes in the four (planar)
directions. The spectrum of the SFAR is given by
\begin{equation}
f(\omega_1,\omega_2) = \frac{1}{4\pi^2 \kappa (1 - 2 \zeta
\cos\omega_1 - 2 \zeta \cos\omega_2)}.
\end{equation}
We define the {\em edge dependence factor} $\zeta$ by
\vspace{-0.5em}
\begin{equation}
\zeta
\defeq \frac{\lambda}{\kappa}, ~~~~ 0 \le \zeta \le 1/4.
\end{equation}
Note that $\zeta =0$ corresponds to the i.i.d. case whereas $\zeta
=1/4$ corresponds to the perfectly correlated case.  Hence, the
correlation strength can be captured in this single quantity $\zeta$ for
 SFAR signals. The power of the SFAR is obtained using the
inverse Fourier transform via the relation (\ref{eq:2DDTFT}), and
is given by
\begin{equation}
P_s = \gamma_{00} = \frac{2K(4\zeta)}{\pi \kappa}, ~~~\left(0 \le
\zeta \le \frac{1}{4} \right),
\end{equation}
where $K(\cdot)$ is the complete elliptic integral of the first
kind \cite{Besag:81JRSS}. The SNR is given by
%\begin{equation} \label{eq:SNR}
$ \mbox{SNR} = \frac{P_s}{\sigma^2} = \frac{2K(4\zeta)}{\pi \kappa
\sigma^2}$.
%\end{equation}
Using  eq. (\ref{eq:errorexponentspectral}) and the SNR, we obtain
the error exponent in the SFAR signal case, denoted by $\Kc_s$ and
given in the following corollary.

\begin{corollary} \label{corol:eeSFA}
The error exponent for the Neyman-Pearson detector for the
hypotheses (\ref{eq:hypothesis2d}) with the SFAR 2D signal model
is given by {\tiny
\begin{eqnarray}
\Kc_s &=& \frac{1}{4\pi^2} \int_{-\pi}^{\pi} \int_{-\pi}^{\pi}
\biggl( \frac{1}{2}\log \left(1+\frac{ \mbox{SNR}}{
(2/\pi)K(4\zeta) (1 - 2 \zeta \cos\omega_1 - 2 \zeta
\cos\omega_2)}\right) \nonumber\\
&& ~~~~~~+\frac{1}{2} \frac{1}{1+\frac{ \mbox{SNR}}{
(2/\pi)K(4\zeta) (1 - 2 \zeta \cos\omega_1 - 2 \zeta
\cos\omega_2)}} -\frac{1}{2} \biggl)d\omega_1d\omega_2.
\label{eq:errorexponentSFA}
\end{eqnarray}
}
\end{corollary}
Note that the SNR and correlation are separated in
(\ref{eq:errorexponentSFA}), which enables us to investigate the
effects of each term separately.

%%%%%%%%%%%%%%%%%%%%%%%%%%%%%%%%%%%%%%%%%%%%%%%%%%%%%%%%%%%%%%%%%%%%
\vspace{-1em}
\subsection{Properties of the Error Exponent $\Kc_s$ }\label{subsec:errorexponent1D}

First,  it is readily seen from Corollary \ref{corol:eeSFA} that
$\Kc_s$ is a continuous function of the edge dependence factor
$\zeta$ ($0 \le \zeta \le 1/4$) for a given SNR. The values of
$\Kc_s$ at the extreme correlations are given by noting that $K(0)
= \frac{\pi}{2} \ \ {\rm and} \ \  K(1)= \infty$.  Therefore, in
the i.i.d. case (i.e., $\zeta =0$), the corollary reduces to
Stein's lemma as expected, and $\Kc_s$ is given by {\scriptsize
\[
\Kc_s= \frac{1}{2} \log (1+ \mbox{SNR}) +\frac{1}{2(1+
\mbox{SNR})} -\frac{1}{2}
=D(\Nc(0,\sigma^2)||\Nc(0,\sigma^2+P_s)).
\]
} For the perfectly correlated case ($\zeta=1/4$), on the other
hand, $\Kc_s =0$. In fact, in this case as well as in the i.i.d.
case, the two-dimensionality is irrelevant. The known result that $P_M \sim \Theta
(n^{-1/2})$ for the perfectly correlated case is applicable.

\begin{figure}[htbp]
\centerline{ \SetLabels
\L(0.25*-0.1) (a) \\
\L(0.72*-0.1) (b) \\
\endSetLabels
\leavevmode
%\ShowGrid
\strut\AffixLabels{
\scalefig{0.22}\epsfbox{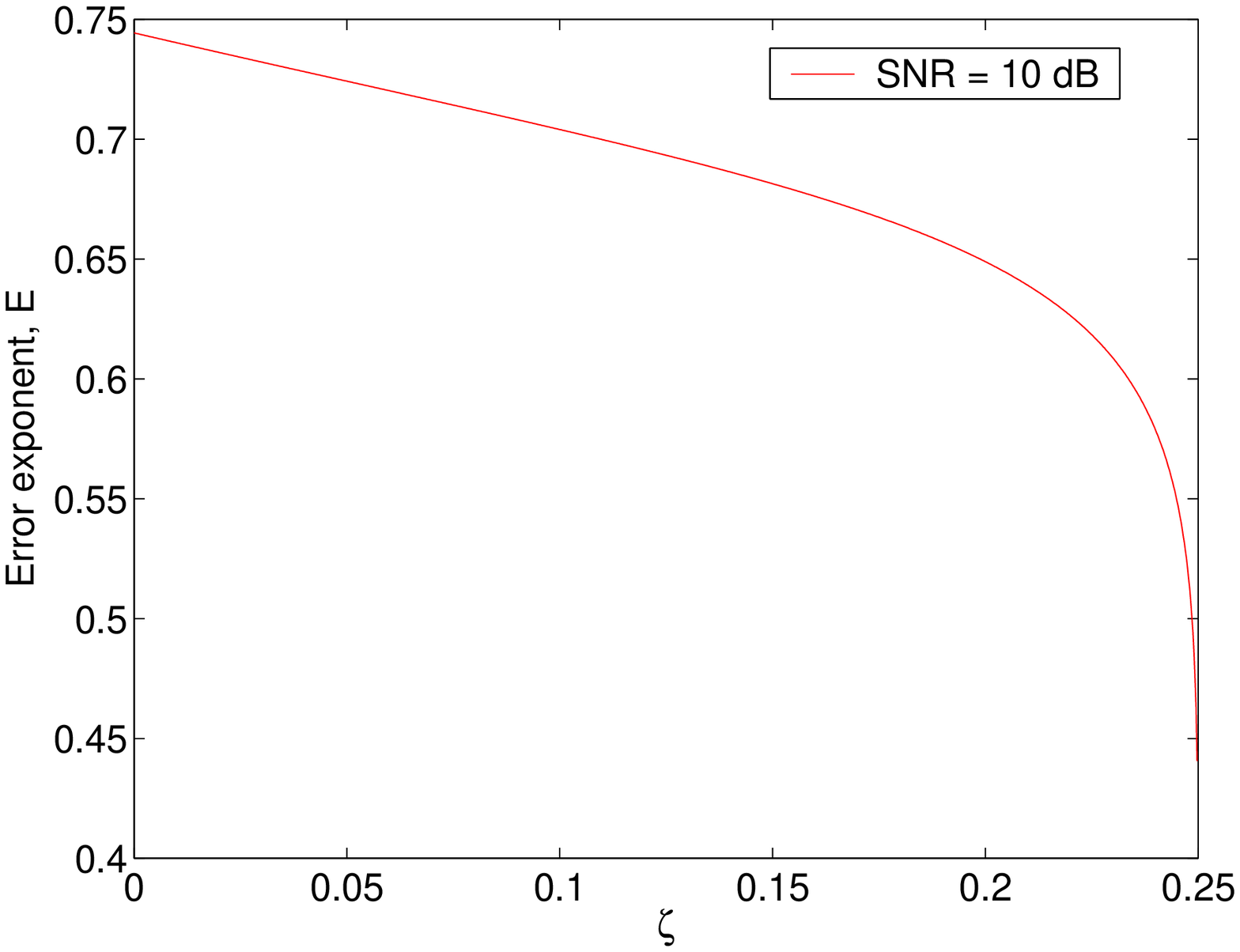}
\scalefig{0.22}\epsfbox{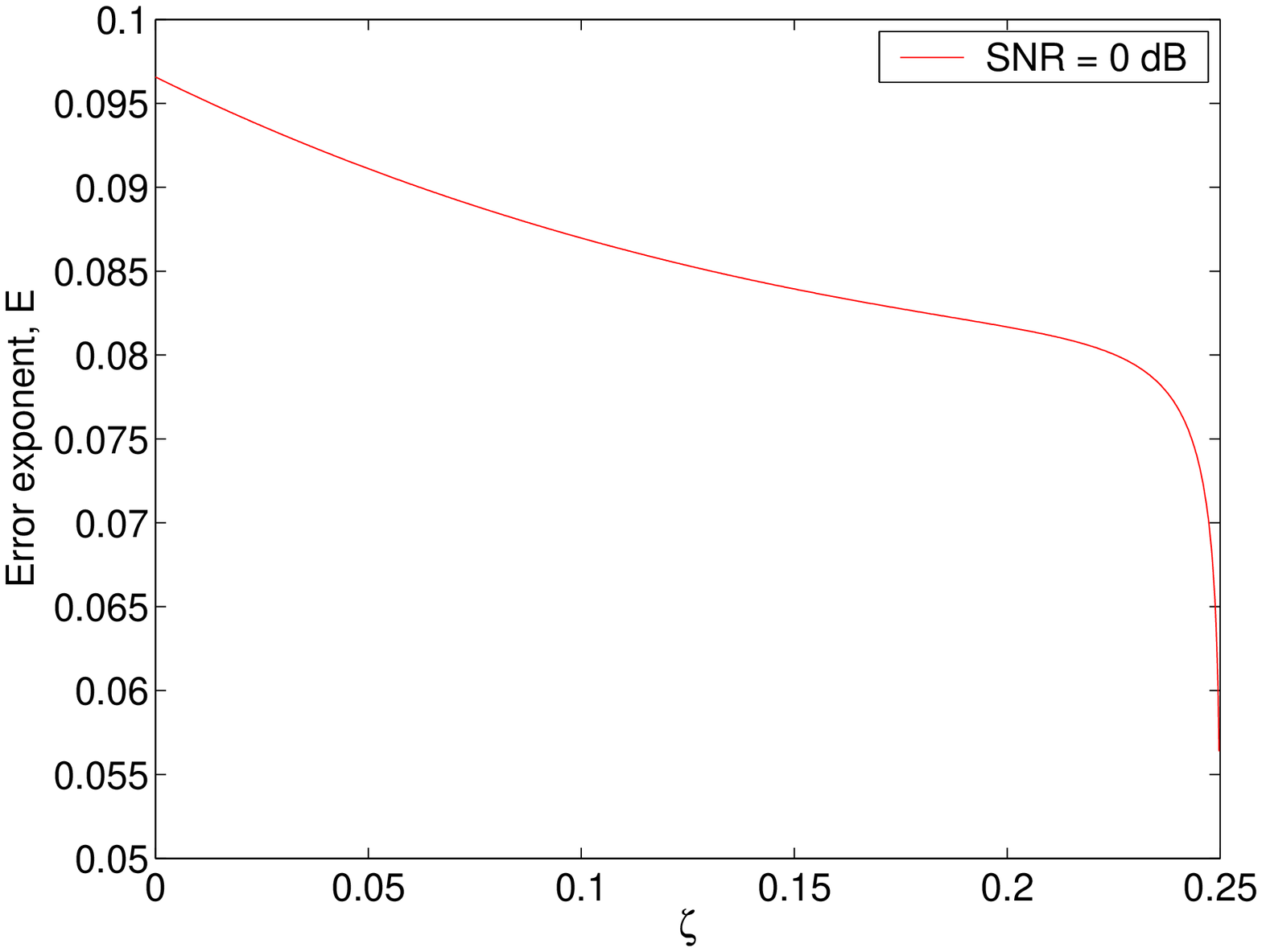}
} } \vspace{0.5cm} \centerline{ \SetLabels
\L(0.25*-0.1) (c) \\
\L(0.72*-0.1) (d) \\
\endSetLabels
\leavevmode
%\ShowGrid
\strut\AffixLabels{
\scalefig{0.22}\epsfbox{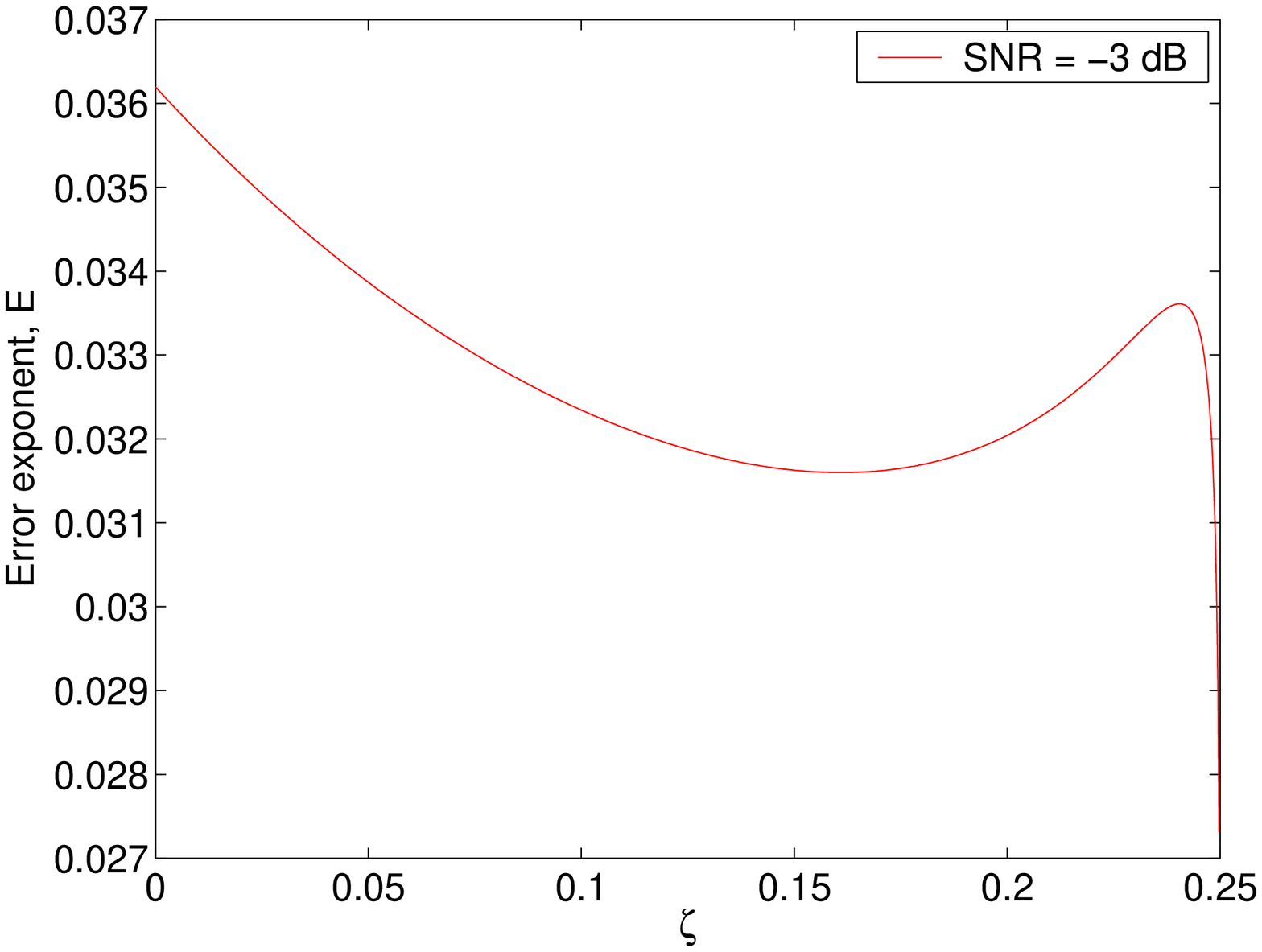}
\scalefig{0.22}\epsfbox{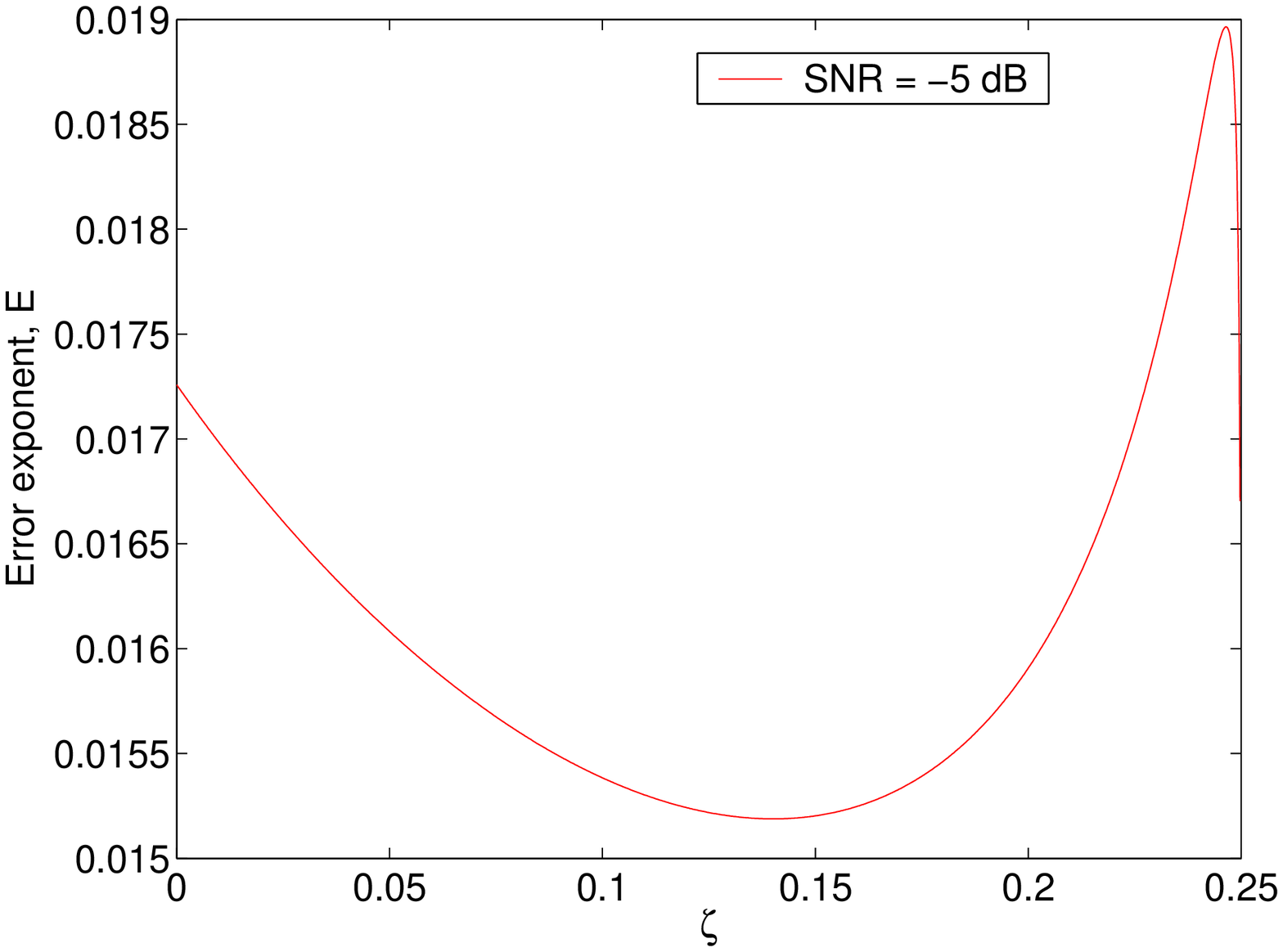}
} } \vspace{0.5cm} \caption{$\Kc_s$ as a function of $\zeta$: (a)
SNR = 10 dB, (b) SNR = 0 dB, (c) SNR = -3 dB, (d) SNR = -5 dB}
\label{fig:KcsVsZeta}
\end{figure}

For intermediate values of correlation, we evaluate
(\ref{eq:errorexponentSFA}) for several different SNR values, as
shown in Fig. \ref{fig:KcsVsZeta}.
It is seen that at high SNR  $\Kc_s$ is monotonically decreasing as
$\zeta$ increases.  Hence,  i.i.d. observations give the best
error performance for  a given value of SNR when SNR is large, as in the 1D
case \cite{Sung&Tong&Poor:06IT}.  As we decrease the SNR, it is observed that a second mode
grows near $\zeta=1/4$.  As we further decrease the SNR, the
value of $\zeta$ of the second mode shifts toward $1/4$, and the
value of the second mode exceeds that of the i.i.d. case. Hence,
there is a discontinuity in the optimal correlation as a function
of SNR in the 2D case even if the maximal $\Kc_s$ itself is
continuous. This is not the case in 1D.

With regard to $\Kc_s$ as a function of SNR, it is straightforward to see
that it is continuous and increases at the rate $\log$ SNR at high SNR for
a given value of $\zeta$.

\section{Ad Hoc Networking: Information-Energy Trade-Off}

\vspace{-1em} The analytical results in the previous section can be
applied to answer some fundamental questions in the design of sensor networks
for detection applications.  We consider a planar {\em ad
hoc} sensor network with minimum hop routing. To simplify the
analysis, we assume that $(2n+1)^2$ sensors are located on the grid $
[-n:1:n]^2$ with spacing $r_n$, as shown in Fig.
\ref{fig:2dHGMRF}, and a fusion center is located at the center $(0,0)$.
We assume that the measurement $Y_{ij}$ is delivered to the fusion center using
the minimum hop routing, which requires a hop count of $|i|+|j|$.

\vspace{-1.25em}
\subsection{Physical correlation model}
\label{subsec:physicalmodel}

The actual physical correlation in this model can be obtained by solving a proper
continuous index 2D stochastic differential equation (SDE), e.g.,
{\footnotesize
\[
\left[ \left( \frac{\partial}{\partial x}\right)^2 +\left(
\frac{\partial}{\partial y} \right)^2 - \xi^2 \right]X(x,y) =
u(x,y),
\]}
where $u(x,y)$ is the process noise and $\xi$ is a parameter
determining the correlation strength of the field. By solving a
proper SDE, the {\em edge correlation factor} $\rho$ is given, as
a function of the edge length $r_n$, by
\[
\rho = f(r_n).
\]
Typically, $f(\cdot)$ is a positive and monotonically decreasing
function of $r_n$.  Further, we have a monotone mapping $g:\rho
\rightarrow \zeta$ from the edge correlation factor $\rho$ to the
edge dependence factor $\zeta$,  which maps zero and one to zero and
1/4, respectively. Thus, we have $\zeta = g(f (r_n)),$ and for
given physical parameters (with a slight abuse of notation),
\[
\Kc_s(\mbox{SNR},\zeta) = \Kc_s(\mbox{SNR},g(f(r_n))) =
\Kc_s(\mbox{SNR},r_n).
\]
We will use the arguments SNR and $\zeta$ for $\Kc_s$ properly if
necessary.

\vspace{-1em}
\subsection{Energy efficiency}
\vspace{-0.5em}

We now consider the energy efficiency of the ad hoc sensor network as
the network size grows. The energy efficiency $\eta$ can be defined as
\begin{equation}
\eta = \frac{{\mbox{total gathered information}~I_t}}{{\mbox{total
required energy}~E_t}},
\end{equation}
where $I_t$ is given by the product of the number of sensors and
the information $\Kc_s$ per each sensor. We consider two
asymptotic regimes for the increase in network size: an infinite area
model with fixed density and an infinite density model with fixed
area. The behavior of the energy efficiency as we increase the
network size is summarized in the following theorems.

\vspace{-0.5em}
\begin{theorem}[Infinite area model] For an ad hoc sensor network
with increasing area and a fixed node density, the energy
efficiency decays to zero as we increase the area with rate
\begin{equation} \label{eq:efficiencyIAM}
\eta = O\left(\mbox{area}^{-1/2} \right).
\end{equation}
\end{theorem}

\vspace{0.3em} {\em Proof:} The total energy required for data
gathering is given by {\footnotesize
\[
E_t = E_{link}(r_n) \sum_{i=-n}^n\sum_{j=-n}^n (|i|+|j|)
=2n(n+1)(2n+1) E_{link}(r_n),
\]}
where the transmission energy per link $E_{link}(r_n) =
r_n^{\delta}$ and $\delta$ is the propagation loss factor.  We
have $I_t = (2n+1)^2 \Kc_s(r_n)$, and $\mbox{area} = \Theta(n^2)$.
The energy efficiency is given by {\footnotesize
\begin{equation} \label{eq:efficientyProof}
\eta = \frac{(2n+1)^2\Kc_s(r_n)}{2n(n+1)(2n+1) E_{link}(r_n)}.
\end{equation}}
  Since $r_n$ is fixed, $\Kc_s$ and $E_{link}$ do not change with
$n$, and (\ref{eq:efficiencyIAM}) follows. $\blacksquare$

\vspace{-0.5em}
\begin{theorem}[Infinite density model]
For the infinite density model, a non-zero efficiency is possible
if the decay rate of the error exponent $\Kc_s$ as a function of
density is slower than
\begin{equation} \label{eq:efficiencyIDM}
O\left(\mbox{density}^{(1-\delta)/2}\right).
\end{equation}
\end{theorem}

\vspace{0.3em} {\em Proof:} For the infinite density model, we
have
\[
r_n = \Theta(n^{-1}), ~~r_n^{\delta}=\Theta(n^{-\delta}), ~~
\mbox{density}=\Theta(n^2).
\]
From (\ref{eq:efficientyProof}), we have $\eta =
\Kc_s(r_n)/n^{1-\delta}$. If $\Kc_s$ as a function of $r_n$ decays
slower than $n^{1-\delta}$, ~$\eta$ does not diminish to zero.
$\blacksquare$

The non-zero efficiency in the asymptotic regime depends on the
decay rate of $\Kc_s$ as a function of $r_n$. Since $\Kc_s(\zeta)$
is given, this depends on the functions $f$ and $g$ in Section
\ref{subsec:physicalmodel} and the propagation loss factor
$\delta$.

\vspace{-1em}
\section{Conclusions}
\label{sec:conclusion} \vspace{-0.5em}

We have considered the detection of 2D GMRFs from noisy
observations.  We have adopted the CAR model for the signal, and
have used a spectral domain approach to derive the error exponent for
 the Neyman-Pearson detector satisfying a fixed level constraint.
 Under the symmetric first order
autoregressive model, we have obtained the error exponent explicitly
 in terms of  the SNR and the edge dependence factor. We have investigated the properties of the error
exponent as a function of SNR and correlation.  We have seen
that the behavior of the
error exponent w.r.t. correlation strength is divided into two
regions depending on SNR.  At high SNR,   i.i.d. (and, thus, uncorrelated) observations
maximize the error exponent for a given SNR, whereas there is
non-zero optimal value of correlation at low SNR. Further, it has been seen that
there is a discontinuity for the optimal correlation as a function
of SNR. Based on the error exponent, we have also investigated the
energy efficiency of {\em ad hoc} sensor network for detection
applications. For a fixed node density, the energy efficiency
decays to zero with rate $O(\mbox{area}^{-1/2})$ as we increase
the area. On the other hand, non-zero efficiency is possible with
increasing density depending on physical correlation strength as a
function of the link length.

%%%%%%%%%% References %%%%%%%%%%%%%%%%%%%%%%%%%%%%%%%%%%%%%%%%%%%%%%%%%%
{\scriptsize
\bibliographystyle{plain}
 }

\end{document}

%% file: macros.tex
\def\psfancypar#1#2{\begingroup\def\par{\endgraf\endgroup\lineskiplimit=0pt}
               \setbox2=\hbox{\large\sc #2}
               \newdimen\tmpht \tmpht \ht2 \advance\tmpht by \baselineskip
               \font\hhuge=Times-Bold at \tmpht
               \setbox1=\hbox{{\hhuge #1}}
               \count7=\tmpht \count8=\ht1
               \divide\count8 by 1000 \divide\count7 by \count8 
               \tmpht=.001\tmpht\multiply\tmpht by \count7 
               \font\hhuge=Times-Bold at \tmpht
               \setbox1=\hbox{{\hhuge #1}}
               \noindent
                \hangindent1.05\wd1
               \hangafter=-2 {\hskip-\hangindent
               \lower1\ht1\hbox{\raise1.0\ht2\copy1}%
                \kern-0\wd1}\copy2\lineskiplimit=-1000pt}

\newcommand{\E}{\mbox{{\rm E}}}

\def\boxit#1{\vbox{\hrule\hbox{\vrule\kern3pt
        \vbox{\kern3pt#1\kern3pt}\kern3pt\vrule}\hrule}}

\def\reals{ { {\rm  I \kern-0.15em R }  } }
\def\complex{ {\,{{\rm C} \kern-0.50em \raise0.20ex {  |}}\, }}

\def\mubf{\hbox{\boldmath$\mu$\unboldmath}}

\def\Sigmabf{\hbox{$\bf \Sigma$}}

\def\ybf{{\bf y}}

\def\ybf{{\bf y}}

\def\Qbf{{\bf Q}}
\def\Rbf{{\bf R}}

\def\Xbf{{\bf X}}

\def\Hc{{\cal H}}

\def\Kc{{\cal K}}

\def\Nc{{\cal N}}

\def\be{\vskip .3cm \begin{equation}}
\def\ee{\end{equation} \vskip .4cm \noindent}
\def\defeq{{\stackrel{\Delta}{=}}}
\newcommand{\R}{\mbox{$\hat {\bf R}_{N}$}}

\def\Rxx{\Rbf_{\ssstyle X\kern-.1em X}}

\let\ssstyle=\scriptscriptstyle

\def\Kout{\setbox1=\hbox{\Huge\bf K}\hbox to
1.05\wd1{\hspace{.05\wd1}
}}
\def\Sout{\setbox1=\hbox{\Huge\bf S}\hbox to 1.05\wd1{\hspace{.05\wd1}
}}

%% file: labelfig.tex
  \ifx\LabelFigloaded\MYundefined\relax
  \else
   \fi

  \def\LabelFigloaded{\relax}% now loaded

  \chardef\LabelFigCatAt\the\catcode`\@
  \catcode`\@=11

 \let\LabelFigwlog@ld\wlog
 \def\wlog#1{\relax}

 \ifx\\\MYundefined@
    \let\\\relax
 \fi

  \def\ms@g{\immediate\write16}

 \def\N@wif{\csname newif\endcsname }
 \def\Temp@ {\N@wif\ifIN@}
 \ifx\INN@\MYundefined@
    \else \let\Temp@\relax
 \fi
 \Temp@

  \def\IN@{\expandafter\INN@\expandafter}
  \long\def\INN@0#1@#2@{\long\def\NI@##1#1##2##3\ENDNI@
    {\ifx\m@rker##2\IN@false\else\IN@true\fi}%
     \expandafter\NI@#2@@#1\m@rker\ENDNI@}
  \def\m@rker{\m@@rker}
 
  \newtoks\Initialtoks@  \newtoks\Terminaltoks@
  \def\SPLIT@{\expandafter\SPLITT@\expandafter}
  \def\SPLITT@0#1@#2@{\def\TTILPS@##1#1##2@{%
     \Initialtoks@{##1}\Terminaltoks@{##2}}\expandafter\TTILPS@#2@}

 \def\Shifted@@#1#2#3{\setbox0=\hbox{#3}%
   \raise -\dp0\vbox {\kern-#2%
       \hbox {\kern#1\unhbox0\kern-#1}%
           \kern#2}}

 \newcount\gridcount
 \newbox\auxGridbox@ \newbox\hGridbox@ \newbox\vGridbox@
 \newbox\Labelbox@ \newbox\auxLabelbox@
 \newbox\Coordinatebox@
 \newtoks\Labeltoks@
 \newdimen\Wdd@ \newdimen\Htt@
 \newdimen\Wddd@ \newdimen\Httt@
 
 \def\Wr@{\immediate\write16}

 \newdimen\GL@wd%% grid-line width
 \def\GridLineWidth#1{\GL@wd=#1}

 \def\gobble#1{}
 \def\EdgeErr@{\Wr@{}%
      \Wr@{\string\Edges\space argument
      1, 10, 100 or 1000 please\string!}%
      }

 \newcount\Edgect@

 \def\Sweepup#1\endSweepup{}

 \def\SetEdges@{%
    \edef\Zr@@s{\expandafter\gobble\number\Edgect@\empty}%
        \count255=0\Zr@@s\relax
        \ifnum\count255=\z@\else\EdgeErr@\show\tailtest\fi
        \count255=1\Zr@@s\relax%\showthe\count255
        \ifnum\count255=\Edgect@\relax\else\EdgeErr@\show\leadtest\fi
    \EdgGl@b\edef\Zr@s{\expandafter\gobble\Zr@@s\empty}%\show\Zr@s
    \ifnum\Edgect@>\@ne\relax\EdgGl@b\let\L@Dc\empty
        \else\EdgGl@b\edef\L@Dc{\string.}\fi
    \ifnum\Edgect@>\@ne\relax
        \EdgGl@b\edef\Edgescale@##1{\divide##1 by \Edgect@}%
        \else\EdgGl@b\edef\Edgescale@##1{}\fi
    }

 \def\Edges#1{\Edgect@=#1\relax
     \let\EdgGl@b\global \SetEdges@}

 \Edges{1}%% default

 \def\hhrule{\hrule height \GL@wd\vskip-.\GL@wd}

 \def\hRule@{%
   \advance\gridcount -2%
   \vfil\hhrule\vfil
   \llap{\smash{\raise -2.5pt
     \hbox{\L@Dc\number\gridcount\Zr@s\kern2pt}}}%
   \hhrule
   }

\def\vvrule{\vrule width \GL@wd \kern-\GL@wd}

 \def\vRule@{\advance\gridcount 2%
   \hfil\vvrule\hfil
   \setbox\auxGridbox@=\vbox to 0pt
      {\vskip \Htt@\vskip 2pt
        \hbox to 0pt{\hss\L@Dc\number\gridcount\Zr@s\hss}\vss}%
      \wd\auxGridbox@=0pt \box\auxGridbox@
   \vvrule
   }

 \def\PlaceGrid@@{\gridcount=10 
  \setbox\hGridbox@=\hbox{%
        \hbox{%
             \hskip-.4pt\vrule
             \vbox to \Htt@{%
               \offinterlineskip\parindent=\z@\relax
               \hbox to \Wdd@{\hfil}
               \hRule@\hRule@\hRule@\hRule@
               \vfil\hhrule\vfil}%
             \vrule\hskip-.4pt}
    }%
  \gridcount=0%
  \setbox\vGridbox@=\hbox{%
      \vbox{\offinterlineskip\parindent=0pt\hsize=0pt
         \vskip-.4pt\hrule%
         \hbox to \Wdd@{%
                 \vtop to \Htt@{\vfil}%
                 \vRule@\vRule@\vRule@\vRule@
                 \hfil\vvrule\hfil}%
         \hrule\vskip-.4pt}}%
  \wd\hGridbox@=0pt\ht\hGridbox@=0pt
  \wd\vGridbox@=0pt\ht\vGridbox@=0pt
  \hbox{\box\hGridbox@\box\vGridbox@}%
  }

 \def\LabelsGlobal{\def\LabGl@b{\global}}
 \def\LabelsLocal{\def\LabGl@b{}}
 \LabelsGlobal %% default

 \def\SetLabels#1\endSetLabels{%
   \LabGl@b\Labeltoks@={#1()\\}%
   }

 \LabGl@b\Labeltoks@={()\\}

 \def\ShowGrid{\LabGl@b\let\PlaceGrid@\PlaceGrid@@}
 \def\HideGrid{\LabGl@b\let\PlaceGrid@\relax}
 \def\Grids{\ShowGrid\LabGl@b\let\GridSwitch@\ShowGrid}
 \def\noGrids{\HideGrid\LabGl@b\let\GridSwitch@\HideGrid}

 \noGrids

 \def\bAdjust@@{%
     \setbox\auxLabelbox@=\hbox{\raise \dp\auxLabelbox@
            \box\auxLabelbox@}}
 \def\bAdjust@{\let\vAdjust@\bAdjust@@}

 \def\eAdjust@@{\dimen0=-.5\ht\auxLabelbox@
     \advance\dimen0 by .5\dp\auxLabelbox@
     \setbox\auxLabelbox@=
            \hbox{\raise\dimen0\box\auxLabelbox@}}
 \def\eAdjust@{\let\vAdjust@\eAdjust@@}

 \def\tAdjust@@{%
     \setbox\auxLabelbox@=\hbox{\raise-\ht\auxLabelbox@
            \box\auxLabelbox@}}
 \def\tAdjust@{\let\vAdjust@\tAdjust@@}

 \let\vAdjust@\relax

 \def\lAdjust@{\let\hAdjust@\rlap}
 \def\rAdjust@{\let\hAdjust@\llap}

 \let\hAdjust@\relax\let\vAdjust@\relax

 \def\FetchLabel@#1(#2)#3\\{%
     \IN@0#2@@\ifIN@
        \setbox0=\hbox{\ignorespaces#1#3\unskip}%
        \ifdim\wd0>0pt
           \ms@g{}%
           \ms@g{ !!! Bad label(s)? !!!}%
           \message{ #1(#2)#3}%
        \fi
        \def\LabelMole@##1\endFetchLabel@{%
            \IN@0()\\@##1@%
            \ifIN@\def\Temp@{\FetchLabel@##1\endFetchLabel@}%
            \else\def\Temp@{}%
            \fi
            \Temp@
           }%
     \else
       \ignorespaces#1\unskip
       \setbox\auxLabelbox@=%
         \hbox to 0pt{\hss\ignorespaces\hAdjust@
          {\ignorespaces#3\unskip}\hss}%
       \vAdjust@
       \let\hAdjust@\relax\let\vAdjust@\relax
       \AugmentLabelBox@@{#2}%
       \ht\Labelbox@=0pt\dp\Labelbox@=0pt
       \let\LabelMole@\FetchLabel@%
     \fi\LabelMole@}

 \newtoks\XYSep@ %\XYSep@{*}
 \def\SetXYSeparator#1{%
     \IN@0#1@@\ifIN@\XYSep@{*}%
     \else
     \XYSep@{#1}%
     \fi
     }

 \SetXYSeparator*

 \def\AugmentLabelBox@@#1{%
     \IN@0\the\XYSep@ @#1@\ifIN@
       \SPLIT@0\the\XYSep@ @#1@%
       \setbox\Labelbox@=\hbox to 0pt{%
         \unhbox\Labelbox@
         \Shifted@@{\the\Initialtoks@\Wddd@}%
         {\the\Terminaltoks@\Httt@}%
         {\box\auxLabelbox@}}%
     \else
         \ms@g{}%
         \ms@g{ !!! Bad insertion point. !!!}%
         \message{ (#1\ this point was rejected.)}%
     \fi
    }

 \def\FetchOption@#1[#2]#3\endFetchOption@{%
    \def\temp{#1}%\show\temp
    \ifx\temp\empty
       \Edgect@=#2\relax%\showthe\Edgect@
       \let\EdgGl@b\relax
       \SetEdges@%\def\Edgescale@##1{\divide##1 by \Edgect@\relax}%
       \Cleaner@#3%
    \fi}

 \def\Cleaner@#1[@]{\Labeltoks@{#1}}
     
 \def\PlaceLabels@@{\mathsurround=0pt%\bgroup
     \def\Cr@{\\}%
     \let\L\lAdjust@\let\R\rAdjust@
     \let\B\bAdjust@\let\E\eAdjust@\let\T\tAdjust@
     \expandafter\FetchOption@\the\Labeltoks@[@]\endFetchOption@
     \Wddd@=\Wdd@ \Edgescale@\Wddd@ %\showthe\Edgect@
     \Httt@=\Htt@ \Edgescale@\Httt@
     \expandafter\FetchLabel@\the\Labeltoks@\endFetchLabel@
     \box\Labelbox@%\egroup
     }%

 \let \PlaceLabels@\PlaceLabels@@

 \def\AffixLabels#1{\setbox\Coordinatebox@=\hbox{#1}%
      \Wdd@=\wd\Coordinatebox@ \Htt@=\ht\Coordinatebox@
      \advance\Htt@ \dp\Coordinatebox@
      \hbox{\copy\Coordinatebox@\kern-\Wdd@ 
           \Shifted@@{0pt}{-\dp\Coordinatebox@}%
           {\PlaceLabels@\PlaceGrid@}%
           \kern\Wdd@}%
      \GridSwitch@ %% next grid hidden
      \LabGl@b\Labeltoks@{()\\}%
      }
 
   \let\wlog\LabelFigwlog@ld   %%restore logging
   \catcode`\@=\LabelFigCatAt  %%12 or 13

                                By

              Raymond S\'eroul <A18645@FRCCSC21.BITNET>
                                and 
              Laurent Siebenmann <lcs@topo.math.u-psud.fr>
    
              VERSIONS: July 1991, Oct 1991, Jan 1992, July 1992

INTRODUCTION

      This labelling package is intended for TeX users who
rely on non-TeX sources for for their graphics inserts.  It
provides means for adding TeX labels to such inserts with a
minimum of fuss. 

       For most labels, TeX users have in the past found it
reasonably convenient to rely on non-TeX sources. Typical
occasions when an inescapable need for TeX labels seemed to
arise are

 (a) when the graphics program lacks certain exotic or complex
mathematical symbols

 (b) when the very highest typographical quality is wanted for the
labels

 (c) when labels included with the graphics fail to print, 
 and you cannot figure out why (cf. boxedeps.doc).  The labels
 provided by labelfig.tex are 100% portable.

       Since this package first appeared, many users, who in the
past scarcely dreamed of using TeX labels, have come to use
nothing but.  So it is now appropriate to add

Intoxication Warning:  TeX labels may be addictive and expensive. 

     If you have a fast preview you may disagree, and even find
that this package provides an agreeable paste-up environment; see
extra applications at end.

     Note to publishers: It is possible and convenient to ultimately
export the TeX labels produced by labelfig.tex to become an integral
part of the EPS file. This is often desired by a publisher who typically
uses an "upmarket" graphics or page layout program, with which the
staff is skilled in perfecting figures.  See Appendix I for
a recipe.

     The authors are grateful to Patrick Ion of Math Reviews for
helpful comments and encouragement.

BASIC INSTRUCTIONS

    After reading in the macro file using

\input labelfig.tex
preview or proof your figure with a coordinate grid printed on
top, by typing the following:

    \ShowGrid  % shows grid  for next figure only
    \AffixLabels{<the graphics insertion>}

Here <the graphics insertion> is what you would type to insert
the graphics object alone without the grid.  This must provide
for the space around it. For example <the graphics insertion>
might well be \BoxedEPSF{MyFigure scaled 700} using the
boxedeps.tex macro package (from same source); this provides a
TeX box containing the encapsulated PostScript insert specified by
the file MyFigure. \AffixLabels{...} provides the grid (supposing
\ShowGrid is present) and later, once you have specified labels
using the grid, it will "tack on" the labels.

     The grid is a sort of (usually elongated) checkerboard of
ten rows and ten columns and its (internal) partitions are by
default numbered  .1, ... ,.9  both horizontally (X-coordinate
running left to right) and vertically (Y-coordinate running bottom
to top).  Thus the points enclosed by the grid correspond to the
points of the unit square in the cartesian "X-Y" plane, the lower
left corner corresponding to the origin (0,0).  By extrapolation,
the full page corresponds to a larger rectangle in the plane.

     These coordinates serve to position labels as follows.
Before the \AffixLabels{...} command type label specifications:

  \SetLabels
   (<X-coordinate>*<Y-coordinate>) <first label> \\
   .
   .
   .
   (<X-coordinate>*<Y-coordinate>)  <last label> \\
  \endSetLabels

Each row specifies one label and is terminated by \\.  In each
row, the position indicator comes first; it is written as a
standard cartesian point except that the X- and Y- coordinates
are separated by * rather than a comma because TeX allows a
comma as decimal point. There are no dimension units to specify
as the unit is the grid itself.

     By default, this cartesian point specifies where the middle
of the baseline of the label will be located.  However if you precede
the point by \L [or \R] the left [or right] edge of the baseline will
be located there. Similarly you may also precede the point by \T, \E,
or \B to vertically align the top equator or bottom of the label box
at the specified point.  This gives nine standard positions of
the label with respect to the insertion point --- corresponding to
the eight principle points of the compas and the center

                     \L\T     \T      \R\T

                     \L\E     \E      \R\E

                     \L\B     \B      \R\B

But this neglects the default "baseline" level of TeX,
giving potentially three more positions

                     \L    <no tag>   \R

For text, the baseline level is often the preferred. Its relation to
the others is variable. It will often coincide with the bottom level,
as happens for "X".  But it is often distinct, as for "g", in which
case you have in all 12 distinct positions rather than 9.

     It is convenient to think of this specification of label
position as attaching the label by a thumb-tack to the coordinate
grid. There are up to twelve positions of the thumb-tack on the
label, while the position of the thumb-tack on the coordinate grid is
arbitrary.  Normally, one choses the position of the thumb-tack on
the label to be the one that is the closest to the item being
labeled.  There are good reasons for this "rule of thumb":

   (a)  It facilitates correct positioning at first try.

   (b)  If the scale of the figure must be altered after labels
have been affixed, the labels have a good chance of remaining well
positioned.

   (c)  The visible grid need not extend beyond the "bounding box"
for the figure, because the best preferred position is always
(at least almost) within the bounding box .

The second reason is particularly important. Indeed it often
happens that scale has to be altered after labelling begins, in
order to either provide space for the labels, or to adjust
proportions between the labels and the figure.  (The size of labels
is unaffected by scaling.)

     Here is an artificial but self-contained test which uses
TeX rules to make a graphics object.

TEST

    Do not skip this!

\input labelfig
 \def\FrameIt#1{\hbox{\vrule$\vcenter {\hrule\kern3pt%
             \hbox {\kern3pt #1\kern3pt}%
               \kern3pt\hrule}$\relax\vrule}}

 \def\Caption#1#2{\FrameIt{%
       \vtop {\hsize=#1\relax \parindent=0pt
         \leftskip=0pt \rightskip=0pt plus15pt
         \parfillskip=0pt
         \lineskip=1pt\baselineskip=0pt
         #2}}}

 \def\FirstQuadrant{\hbox to 100pt{\vrule\vbox to 100pt{%
        \hbox to 100pt{\hfil}\vfil\hrule}\hss}}

  \SetLabels
    \R(.5*.2) $\zeta\,\cdot$\\
    (.9*-.10) $\xi$\\
    \R(-.03*.9) $\eta$\\
    \T(.5*.9) \Caption{70pt}{%
          \it The norm of
          $g(\xi+i\eta)$ is indicated on
          contours of this invisible surface.}\\
  \endSetLabels

  \AffixLabels{\FirstQuadrant}

  \end

  Note that the coordinates to use for labels are indicated on the
edges of the grid (when visible) corresponding to the conventional
x- and y- axes of the Cartesian plane. By default the grid is
1-by-1. However, by the command \Edges{100}, you can change this
to 100-by-100 and many users find this alternative most
convenient. Place the command \Edges{...} in your style file (or
header) since its effect is is global. Other possible edge values
are 10 and 1000.

  If you use the command \Edges{...} at all, do so with care.  For
if you accidentally delete an \Edges{...} command your labels will
abruptly be badly misplaced and may logically but mysteriously
generate "dimension too big" errors under TeX and "off page" errors
under your driver.  

  You can dictate the edgescale for an individual figure by giving
the scale in brackets immediately after \SetLabels.  Thus, to
import into an article using say \Edge{100} a figure labelled using
another edgescale, say the original 1-by-1 default, you can use
\SetLabels[1]...\endSetLabels.

GETTING IT DOWN PAT

     Complicated labeling deserves the same respect as
complicated mathematics.  Do not expect it to come out perfect the
first time!  What is needed in either case is a mechanism to
repeatedly typeset troublesome pieces.

     One mechanism is always available.  One does complicated
labelling in a separate "test" file involving just the figure being
labelled;  a texpert will know how to \dump TeX's current state as
a temporary format that restarts rapidly at each retry.  Usually,
one then pastes the completed labelled figure back into the main
TeX file, but, of course, one can also \input it as an auxiliary
file.

     If you do not have a TeXpert at handy, here is a first
approximation to an efficient setup. By deletions reduce a copy
of your article to just a few lines before and after the figure.
Now label the figure, and finally, copy and paste the labelled
figure to the original article. Then copy the next figure to label
into this testbed and repeat. The TeXpert can improve the  speed
at which TeX starts up, by compiling a format specifically for
your article; just one caution: best NOT include in the format
ephemeral details of setup like \Set<mydriver>ArtSpecials (from
boxedeps.tex because this reads  figure dimensions which you may
change during your work session.

     An improved mechanism to repeatedly typeset troublesome
pieces is now available on the Macintosh; it is called LinoTeX;
see the same ftp sources.  It could be set up on many types
of computer.

     Before using labelfig.tex to attach labels to a graphics
object inserted using boxedeps.tex or BoxedArt.tex, make it a
firm rule to carefully adjust the bounding box using the trimming
commands of these packages, and also at least tentatively scale
and position the object. Beware of changing the grid inadvertently
after the labels have been positioned.  For example, correcting
the bounding box of a PostScript graphics object can foul up the
labels by changing the coordinate grid to which the labels are
attached. This is particularly true for the trimming  commands of
boxedeps.tex and BoxedArt.tex. However, as noted already, change
of scale is much less disruptive, and modest adjustments should be
well tolerated.

     Sometimes the labels protrude so far from the bounding box
of a figure that the figure has to be repositioned.  Best do this
by ad hoc spacing, say using \hglue and \vglue; altering the
bounding box would create a vicious circle.

     Remember that you are responsible for preventing labels
from overlapping. You are responsible for all label typography
including size and style. A label is really just about anything
that can be put in a TeX box. Note that spaces at the beginning
and end of labels will normally be suppressed; if you really want
them you must protect them with TeX braces.

     This package temporarily sets the \mathsurround parameter
of TeX to zero  while the labels are being affixed. This is done
because nonzero \mathsurround space would influence the position
of left and right aligned labels; then, when a texpert or printer
modifies mathsurround, diagram labeling might be disastrously
altered. There is a small price to pay involving labels that are
formatted as caption boxes including mathematics: you  may want or
need to specify an explicit mathsurround space within the caption
box; it will not influence anything outside.

     Those hostile to the use of * as separator between
the X and Y coordinates of label insertion points, are free to
impose another using \SetXYSeparator{<the new separator>}.  
Americans may prefer "," to "*" since they never use a 
comma as a decimal point; on the other hand, * may be more visible.

APPENDIX (I)  MERGING labelfig.tex LABELS INTO AN EPSF GRAPHICS OBJECT.

     As promised in the introduction, here is a recipe useful for
publishers. It works at least on Macintosh and at least for vectorized
graphics and Adobe type1 fonts.  (There is surely a similar recipe for
PCs under MSWindows.)

 (a)  Use boxedeps.tex utility to integrate the figure given by the eps
file, "x.eps" say, with a visible frame around it.  See
\ShowDisplacementBoxes command in boxedeps.tex.  To get precise results
automatically it is important to use the \Trim... commands of
boxedeps.tex making the "DisplacementBox" neatly fit the figure.

 (b)  Use the TeX printer driver and LaserWriter (versions >= 8.1.1) to
export to an EPSF the DVI page containing the integrated, labelled
figure. You now have an EPS file  "xx.eps"  that contains too much, and at
the wrong scale, and at wrong position.

 (c)  Convert the EPSF to an Adode Illustrator format EPSF using
the shareware utility called epsConvert by Sam Weiss
1993-- (currently $25).

 (d)  In Illustrator (or a compatible program), group the labels and the
"DisplacementBox"; copy them to the clipboard and paste them into "x.ps".
This step requires that all the label fonts be "visible to the Macintosh.

 (e)  Translate and scale the pasted group consisting of the labels plus
the "DisplacementBox" so as to make the "DisplacementBox" the bounding
box of (labelless) figure represented by "x.eps".  At this point the
labels will be correctly placed on the figure "x.eps".

 (f)  Ungroup and delete the "DisplacementBox".  The result is the
desired single EPS file, "x+.eps" say, It contains the original figure
plus its labels.  

     Using grouping and ungrouping appropriately in "x+.eps", a
publisher's staff can very efficiently improve label positions etc.

APPENDIX II)  SOME EXOTIC APPLICATIONS

     The grid of labelfig.tex is analogous to a light-table in
classical page makeup with wax or latex glue.  In principle, you
can use it to compose any page from its indivisible parts.  This
even has some of the artisanal charm of classical paste-up
provided you have a fast screen preview to make the process
"interactive".

     In practice labelfig.tex is a tool for nonstandard jobs.
Here are a few going beyond the labelling already discussed.

(I)  GRAPHICS INTEGRATION.

     This is accomplished by treating the imported graphics
objects as labels.  The underlying graphics object is then
typically an empty  \vbox to <dimension>{\vfill} in a TeX
\midinsert...\endinsert construction.  A label line
might be of the form

   (.1*.1) \\

The exact form of the special command varies from driver to
driver.  However, in the case of encapsulated PostScript graphics
(EPSF norm), by relying on boxedeps.tex, one can have the
following standard syntax (independant of driver  (see
boxedeps.doc for details.
  
  (.1*.1) \BoxedEPSF{MyFigure scaled <scale in mils>}\\

This may be slow since it requires TeX to read the PostScript
file to read bounding box using many complex macros.  So you
may want to try

  (.1*.1) \EPSFSpecial{MyFigure}{<scale in mils>}\\

which is fast and driver independant, but it squashes the
bounding box, normally to its lower left corner.

     Similarly for graphics of the Macintosh PICT norm ---
using BoxedArt.tex (same sources) in place of boxedeps.tex.

     This approach to integration is to be recommended when
one is assembling a composite graphics object.

 (II)  COMMUTATIVE DIAGRAM ENHANCEMENT

     Commutative diagrams or arrays of mathematical objects
connected by arrows of various sorts are common in mathematics.
The mathematical objects require the use of TeX.  Recently TeX
acquired a good collection of arrows of all slopes --- that of
LamSTeX --- plus pwerful macros to build the diagrams.

     However, even the LamSTeX collection is often
inadequate; it lacks for example double shafted arrows, dotted
arrows and curved arrows. Fortunately it is possible to produce
such arrows on an individual basis using sophisticated graphics
programs such as Illustrator and AldusFreehand (both serving
the EPSF norm) or using Metafont (with its public domain norm).
Since the creation of each new arrow is a work of love, you
probably want to limit the number of arrows by using LamSTeX
for most arrows. The 40K commutative diagram module of LamSTeX
has been adapted to work with AmSTeX and a copy may be posted
with LabelFig and related files. Unfortunately no one has yet
offered a version that works with Plain TeX or LaTeX.

       Suffice it here to say that when the exotic arrow has
been somehow imported into TeX, labelfig.tex treats it as a
label that one affixes to the commutative diagram.  Two other
steps will be treated in separate notes, namely the matter of
extracting the dimension specifications for the arrow and the
construction of the arrow --- for these steps are far from
unique and often depend intimately on your computer environment. 
Notes for the Macintosh-Textures-Illustrator combination are
found in the file ExoticArrows.doc.

 (III) NESTING 

Ingenuity pays off in exploiting labelfig.tex. One can
mix graphics and typography quite freely.  labelfig.tex is good
for freeform or overlapping arrangements, while boxedeps.tex (or
BoxedArt.tex) is best for regimented non-overlapping
arrangements --- and the two can be combined.

     The default behavior of labelfig.tex is not ideal 
for nesting objects, because to prevent trouble for beginners
the register for labels is globally cleared when \AffixLabels
concludes.  But there are switches available

      \LabelsGlobal      \LabelsLocal

which change this.  To understand this, extend the above test 
by something like:

 \LabelsLocal
 \SetLabels
    (.5*.5) AAA\\
 \endSetLabels

 {%%% Watch for influence of braces!!
 \SetLabels
    (.5*.5) ZZZ\\
 \endSetLabels
   \AffixLabels{\FirstQuadrant}
 }

   \AffixLabels{\FirstQuadrant}

     There are however potential pitfalls.  Neither
labelfig.tex nor boxedeps.tex has been tested under extreme
conditions. Problems may occur if their procedures are
indiscriminately nested. For boxedeps.tex (not labelfig.tex)
there is a precise cause for worry, namely many of its
variables are "global", which means that TeX braces will not
provide the protection one might expect.

COMMAND SUMMARY FOR labelfig.tex

  Here [...] means optional (one or zero)
       [...]* means any number of such constructs

  \SetLabels
    [[<P>](<X><Sep><Y>) <label> \\]*
  \endSetLabels
  \ShowGrid  % this makes the grid visible (once)
  \AffixLabels{<the figure>}

   --- <P> is tack position, one of eleven or empty
              order irrelevant

                   \L\T      \T      \R\T

                   \L\E      \E      \R\E

                     \L               \R

                   \L\B      \B      \R\B

   --- (<X><Sep><Y>) insertion point;
  <Sep> is separator, = * by default;
  \SetXYSeparator{<Sep>} changes it.
   <X> and <Y> are real numbers

  --- <label> a label to attach 

  --- <the figure> the figure to label 

  \GlobalLabels (default)     
  \LocalLabels  setting for nested constructs.

 \Grids makes ALL grids appear; \HideGrid then makes just next disappear.
 \noGrids returns to default.  The commands are always global.

 \GridLineWidth{<dimension>} adjusts width of grid lines. Default is very
small, to give "hairline" effect. If your grid lines are missing try
setting \GridLineWidth{1pt}.

 \Edges#1 globally changes the edge size of all grids to the numerical 
value #1, which must be 1, 10, 100, or 1000.  The default is 1.

VERSION HISTORY.
 --- Jan 1993: \Edges#1 and [??] option after \SetLabels
 --- July 1992: \Grids, \noGrids, \HideGrid;
       Gridlines become hairlines; \GridLineWidth{<dimension>}.
 --- Oct 1991, Jan 1992: \SetXYSeparator{<Sep>},  \LabelsGlobal,
       \LabelsLocal.
 --- July 1991: first release

Address for bugs and other feedback:

        Raymond S\'eroul
        IREM and Lab. de Typographie Informatise
        Univ. Rene Descartes
        Strasbourg

    Tel 33-88-41-63-45
    Email:  A18645@FRCCSC21.BITNET

        Laurent Siebenmann
        Mathematique, Bat. 425,
        Univ de Paris-Sud,
        91405-Orsay,
        France

    Tel 33-1-6941-7949; 
    Email: lcs@topo.math.u-psud.fr